\documentstyle[aps,epsf,twocolumn]{revtex}

\def\Lsu2{{\cal L}_{\mbox{SU(2)}}}

\def\su2xsu2{{SU(2)\times SU(2)}}
\def\su3xsu3{{SU(3)\times SU(3)}}

\def\ben{\begin{eqnarray}}
\def\een{\end{eqnarray}}

\newcommand{\be}{\begin{eqnarray}}
\newcommand{\ee}{\end{eqnarray}}

\newcommand{\beq}{\begin{equation}}
\newcommand{\eeq}{\end{equation}}
\newcommand{\bea}{\begin{eqnarray}}
\newcommand{\eea}{\end{eqnarray}}

\begin{document}

\draft
\title{\bf  Lattice QCD Spectra at Finite Temperature :\\
                    a Random Matrix Approach}

\author{{\bf Maciej A.  Nowak}$^{1}$, {\bf G\'abor Papp}$^{2}$ 
and {\bf Ismail Zahed}$^3$}

\address{$^1$
GSI, Plankstr.1, D-64291 Darmstadt \& Institut f\"{u}r Kernphysik,
TH Darmstadt, Germany \& \\Department of Physics,
Jagellonian University, 30-059 Krakow, Poland;\\
$^2$GSI, Plankstr. 1, D-64291 Darmstadt, Germany \&\\
Institute for Theoretical Physics, E\"{o}tv\"{o}s University,
Budapest, Hungary\\
$^3$Department of Physics, SUNY, Stony Brook, New York 11794, USA.}
\date{\today}
\maketitle

\begin{abstract}
We suggest that the lattice Dirac spectra in QCD at finite temperature may be
understood using a gaussian unitary  ensemble for Wilson 
fermions, and a chiral gaussian  unitary ensemble for 
Kogut-Susskind fermions. For  Kogut-Susskind fermions, the lattice results 
by the Columbia group are in good agreement with the spectral distribution 
following from a cubic equation, both for the valence quark distribution and 
the anomalous symmetry breaking. We explicitly construct a number of Dirac 
spectra for Wilson fermions at finite temperature, 
and use the end-point singularities to derive analytically
the pertinent critical lines. For the physical current masses, the matrix model
shows a transition from a delocalized phase at low temperature, to a localized
phase at high temperature. The localization is over the thermal wavelength of 
the quark modes. For heavier masses, the spectral distribution reflects on
localized states with competitive effects between the quark Compton wavelength 
and the thermal wavelength. Some further suggestions for lattice simulations 
are made.

\end{abstract}

{\bf 1.\,\,\,}Numerous lattice simulations indicate that massive QCD 
undergoes either a rapid cross-over or a phase transition at high 
temperature. The character and nature of these phase changes are still 
debated \cite{KANAYA}. Most numerical analyses
on the lattice have relied on Kogut-Susskind (staggered) quarks, although 
results using standard or modified Wilson actions are now becoming available 
\cite{IWASAKI}.

QCD with three flavors undergoes a first order transition for light current 
quark masses. The transition is believed to switch to a second order for zero
up and down quark masses with a heavy strange quark, and a cross-over for 
all other finite quark masses.
When the quarks become very heavy, they decouple. A pure Yang-Mills phase
undergoes a first order transition at high temperature. The precise values of 
the critical temperature, and critical quark masses are not yet unambiguous, as 
they seem to depend on scaling and the way the quarks are set on the lattice 
\cite{DETAR}.

The lattice results have spurred a number of theoretical investigations
using mostly effective models, all  aimed at understanding the whys
behind the disappearance of chiral symmetry whether through a sharp phase
transition or a smooth cross-over \cite{BOOK}. Most noteworthy are the
suggestions made by Pisarski and Wilczek concerning the generic character of 
the chiral phase transition and its relation to scaling and universality 
\cite{PISARSKI}.

The purpose of this letter is to use some insights from random matrix theory
\cite{RANDOMOTHERS},
to analyze some aspects of chiral symmetry in QCD at finite temperature and 
nonzero current quark masses \cite{RANDOMQCD,BEFORE,STEPHANOV}. In section 2, we show 
that the recent finite temperature 
results by the Columbia group \cite{CHRIST}
using two-flavor QCD with staggered fermions,
in the semi-quenched approximation are well described by a high
temperature version of the chiral gaussian unitary ensemble with thermal 
masses, as expected from dimensional reduction. The spectral distribution 
follows from a cubic (Cardano) equation. The issue related to the 
lattice simulation of the anomalous symmetry breaking is also recovered, 
although the chiral random matrix model does not account for the U(1) 
anomaly. In section 3, we suggest that three-flavor QCD with Wilson fermions,
in the quenched approximation, follows from a gaussian unitary ensemble.
The role of the various Matsubara modes is highlighted. In section 4,
the Dirac spectra for different masses and temperature are just given by 
a superposition of three solutions to a cubic (Cardano) equation. The phase
diagram of the random matrix model is shown to follow from the end-point
of the Dirac spectrum analytically. Our conclusions and recommendations are 
given in section 5.

\vskip .5cm
{\bf 2.\,\,\,}
Recent lattice simulations by the Columbia group~\cite{CHRIST} using staggered 
fermions have unraveled interesting aspects of the Dirac spectrum of 
two-flavor QCD in the semi-quenched approximation on a $16^3\times 4$ and 
$32^3\times 4$ lattices. If we were to denote by $\zeta$ the valence quarks 
with mass $m_{\zeta}$, then the valence quark condensate \cite{CHRIST}
\footnote{$\varrho(\lambda 
)\sim |\lambda |^3$ asymptotically in QCD, and (\ref{chris1}) diverges 
quadratically. The finite lattice spacing provides a natural
ultraviolet cutoff, making (\ref{chris1}) finite. A better definition of 
(\ref{chris1}) is on a cooled lattice. In random matrix models, the spectra
are bounded.}, 
\be
<\overline{\zeta}\zeta>(m_{\zeta}) = 2m_{\zeta} \int_0^{\infty}d\lambda\ 
\frac{\varrho(\lambda)}{\lambda^2+m_{\zeta}^2}
\label{chris1}
\ee
where $\varrho (\lambda)$ is the Dirac 
eigenvalue distributions associated to $iD\!\!\!\!/$ 
for two-flavor and massive QCD. The third flavor $\zeta$ is not included while
averaging over the sea fermions. The dependence on the sea mass $m_s$ (not to 
be confused with strangeness) and the number of flavors in 
$\varrho (\lambda )$, stem from the random averaging over the 
gauge configurations in the presence of the two-flavor fermion determinant 
\cite{CHRIST}. For two degenerate flavors with $m_sa=0.01$,  the behavior of 
(\ref{chris1})  versus $m_{\zeta}$ on a $16^3\times 4$ lattice 
is shown in Fig. 1, for eight temperatures $\beta =6/g^2$ around the critical 
temperature $\beta_c= 5.275$. The transition in this case is believed to be 
second order, with a critical temperature $T_c=1/4a\sim 150$ MeV \cite{DETAR}. 
In physical units, the lattice spacing at the critical temperature is $a\sim 
.33$ fm, and the sea quark mass is $m_s=6$ MeV.

\begin{figure}[tbp]
\centerline{\epsfxsize=7.5cm \epsfbox{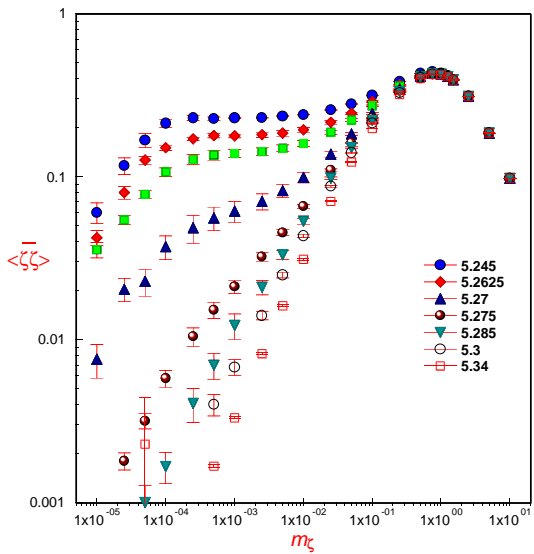}}
\caption{Semi-quenched condensate versus the valence quark mass $m_{\zeta}$
for two-flavor QCD \protect\cite{CHRIST} with a sea mass $m_sa =0.01$.}
\label{fig1}
\end{figure}

The bulk of the staggered lattice results can be  understood from simple random 
matrix theory \cite{RANDOMQCD,BEFORE,STEPHANOV}, if we note that for a bounded spectral density, the valence 
condensate is just
\be
<\overline{\zeta}\zeta>(m_{\zeta}) = \pi \varrho(i m_{\zeta}) .
\label{chrisa1}
\ee
which is just the spectral density for $\lambda =im_{\zeta}$. For staggered 
QCD fermions at zero temperature, the spectral distribution can be mapped on 
a chiral gaussian unitary ensemble, provided that the gauge configurations are 
sufficiently random. At finite temperature, this is equivalent to an assembly 
of Matsubara modes, each described by a gaussian unitary ensemble. At high
temperature the only relevant modes are $\pm \pi T$, as suggested by the 
screening length measurements \cite{DETAR}, and dimensional reduction
\cite{REDUCTION}. We note that $\pi T$ is the minimum wavenumber 
$k=2\pi/\lambda$ of a massless quark mode on a circle of length $1/T$
with a wavelength $\lambda = 2/T$. For massive quarks, the relevant modes 
are $\pm M(T) = \pm\sqrt{m_s^2 +\pi^2 T^2}$ \cite{REDUCTION}. With this in mind, a
suggestive high temperature chiral random matrix model for two degenerate
and massive flavors is
\be
{\bf Q}_S = \bigotimes_{\pm; f=1,2} \left(
\left( \begin{array}{cc} 0 & \pm M(T) \\
                              \pm M(T)  & 0 \end{array}
                       \right)  +
            \left( \begin{array}{cc} 0 &  R\\
                                    R^{\dagger} & 0\end{array} \right) \right)
\label{1}
\ee
where each entry in (\ref{1}) is $N\times N$ valued. 
The $R$'s are described by 
a gaussian unitary ensemble. The off-block-diagonal structure of (\ref{1}) is
suggestive of the chiral structure of $iD\!\!\!\!/$, which is preserved in the 
staggered fermion formulation. 
This is not the case for the Wilson fermions, as 
we discuss below. All the scales in (\ref{1}) are given in units of 
$1/\Sigma=(100 \,\,{\rm MeV})^{-1}$. Although (\ref{1}) is well-motivated  at high temperature, 
we will use it to investigate the temperature ranges around the critical 
temperature. This is a bit justified by the fact that the lattice simulations
of the spatial screening lengths show a behavior that is consistent with the
occurrence of the lowest Matsubara modes  around $T\sim T_c$. For $m_s=0$, 
and in the case of one-flavor and a single Matsubara mode,
(\ref{1}) has been numerically investigated in \cite{BEFORE}.

The spectral distribution associated to (\ref{1}) follows from the 
discontinuity of the pertinent solution to a cubic (Cardano) equation,
\be
G^3(z)-2zG^2(z)+(z^2- M^2(T) +1)G-z=0
\label{CAR}
\ee
with $-\pi\varrho (\lambda ) = {\rm Im } G (\lambda + i 0 )$, this was also
observed in \cite{STEPHANOV} for the massless case. The equation (\ref{CAR})
follows readily from the law of addition of random plus deterministic matrices
\cite{US}.

\begin{figure}[tbp]
\centerline{\epsfxsize=7.5cm \epsfbox{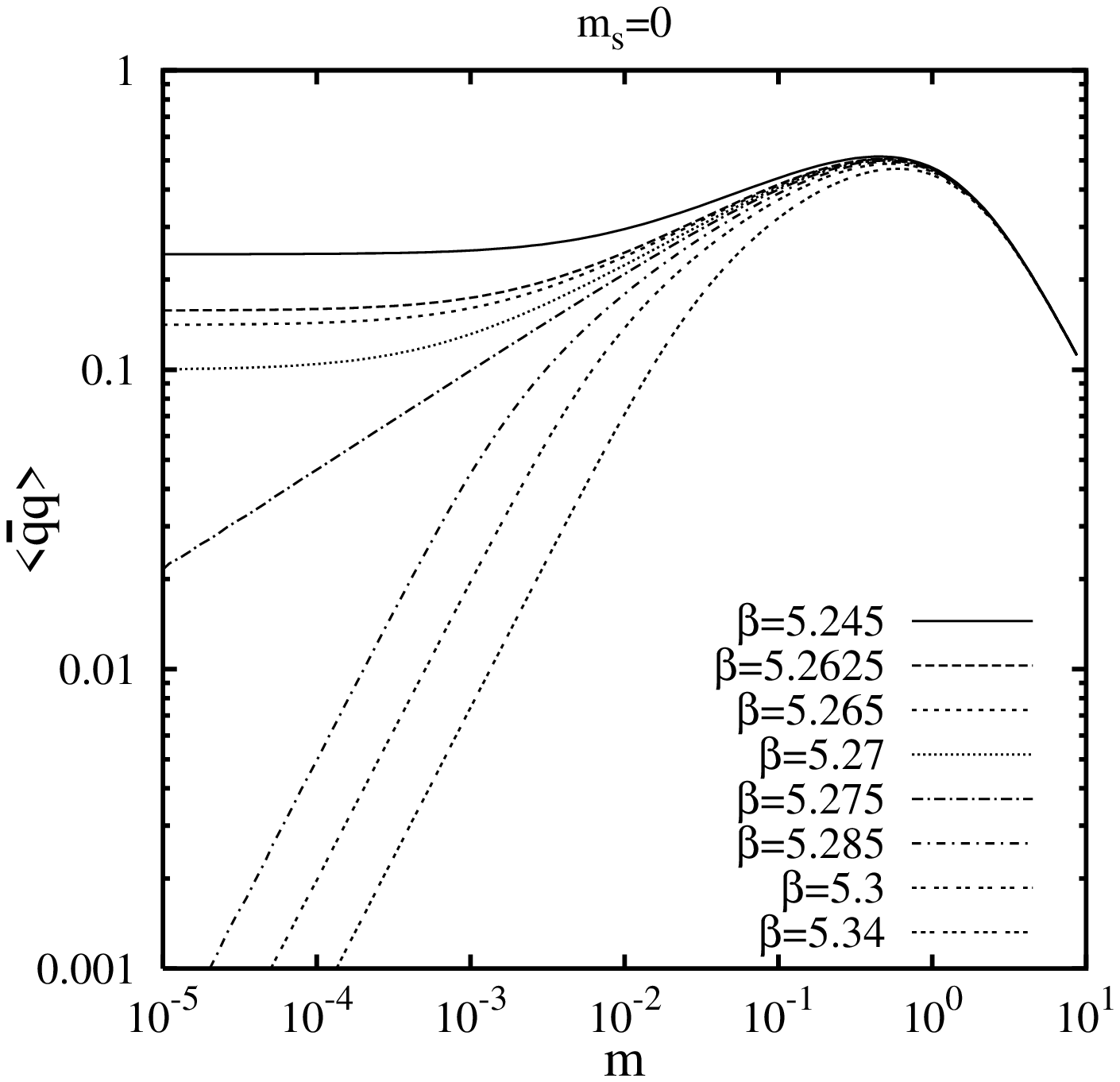}}
\centerline{\epsfxsize=7.5cm \epsfbox{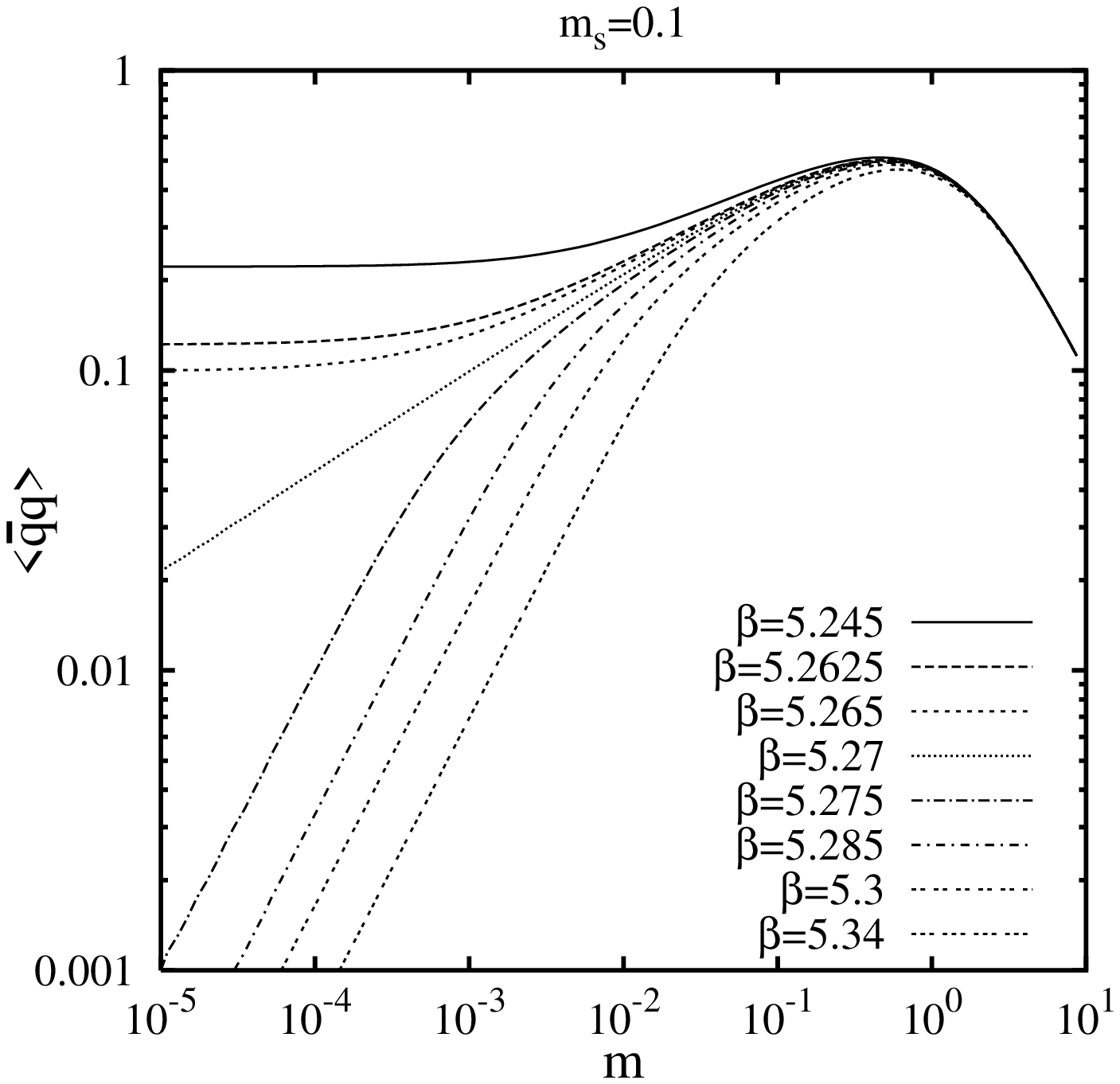}}
\caption{Semi-quenched condensate versus the valence quark mass $m_{\zeta}$
for two-flavor QCD, obtained from eq.(\protect\ref{chrisa1}) with 
the Cardano solution of eq.(\protect\ref{CAR}),
for the sea quark masses $m_s=0$ (upper) and $m_s=0.1$ (lower).}
\label{fig2}
\end{figure}

In Fig.~\ref{fig2},
we show the behaviors of the valence quark condensate for two choices of the 
sea quark masses, $m_s=0$ and $0.1$, and  various $\beta$ 
(temperature) values as obtained from the Cardano solution (\ref{CAR}).
In physical units, $m=0$ and $10$  MeV.  We have identified $\beta$ with $T$, 
through\footnote{Note that here $\beta$  is explicitly related to
the temperature, while on the lattice $\beta=6/g^2$ is implicitly related
to the temperature through the coupling constant after a change in the lattice 
spacing. The use of the same parameter for both denominations is only 
suggestive.}
\be
\pi ( T-T_c) = (\beta -\beta_c)
\label{CRI}
\ee
with the critical temperature $T_c$ in the chiral random matrix model and 
$\beta_c= 5.275$ as suggested by the lattice calculations \cite{CHRIST}. 
The transition in the chiral random matrix model is mean-field in character
(large $N$). For $m_s=0.1$ (10 MeV) the random matrix model seems to follow 
qualitatively well the lattice results for $m_sa=0.01$ (6 MeV), except for 
$m_{\zeta}\leq 10^{-5}$, where the spectrum becomes sensitive to the finite 
size of the lattice, as illustrated by the bending of the upper curves. Finite
size effects set in when the pion Compton wavelength becomes comparable to the 
lattice size. Using the PCAC relation this implies that 
$m_{\zeta}<\overline{\zeta} \zeta> \geq 1/N_cV_4$. For $V_4=4\times 16^3$, this
puts a lower bound on the valence mass $m_{\zeta} \geq 10^{-4}$, which is about
consistent with the lattice results. We note that the above interpretation of 
the finite temperature lattice simulation is different from the zero 
temperature universality  argument used in \cite{MICRO}, but
in qualitative agreement with the finite-N numerical analysis carried
in \cite{BEFORE}, in the massless case.

We note 
that the reading of the critical temperature $\beta_c$ depends sensitively on 
the value of the sea quark mass, as can be seen by comparing the upper and 
lower curves on Fig.~\ref{fig2}. For $m_s=0$, the critical temperature
is $\beta=5.275$ (fifth line from the top), whereas for 
$m_s=0.1$, the critical value is $\beta=5.27$ (the fourth line
from the top).  In the random matrix model (Cardano solution), the dependence 
of the critical temperature on the sea quark mass is simply 
$\Delta \beta_c =\sqrt{1-m_s^2}$.  Both lines have the same critical exponent 
$\delta=3$ (mean-field). This result was first established in \cite{BEFORE}.
For $\beta=5.275$, $<{\overline\zeta} \zeta >
\sim  m_{\zeta}^{0.33}$. For $\beta= 5.34$, $<{\overline\zeta} \zeta > = 8 
m_{\zeta}$. These values are to be contrasted with the ones quoted by 
Chandrasekharan and Christ \cite{CHRIST}, namely 
$<{\overline\zeta} \zeta >  \sim  m_{\zeta}^{0.6}$ for
$\beta = 5.275$, and $<{\overline\zeta} \zeta >
\sim  m_{\zeta}^{0.98}$ for $\beta =5.34$. The closeness of the 
latter to the random matrix result, suggests perhaps an underestimation of the 
critical exponent $\delta = 5/3$ \cite{CHRIST} in comparison with the 
mean-field one $\delta= 3$, possibly due to the large finite sizes effects 
mentioned above as visible in Fig.~\ref{fig1} (dropping plateau's for small 
masses). We recall that a two-flavor simulation of QCD using 
finite-temperature cumulants yields $\delta$
in the range 4-5 \cite{KARSCH}, hence closer to the mean-field result.

The fact that the semi-quenched condensate appears to vanish linearly with the 
valence quark mass $m_{\zeta}$ in the high temperature phase ($\beta > 
\beta_c$) prompted the Columbia group to analyze the asymmetry \cite{CHRIST}
\be
\omega = 4 m^2 \int_0^{\infty} d\lambda \ 
\frac{\varrho(\lambda)}{(\lambda^2+m^2)^2}
\label{chris2}
\ee
between the isotriplet and its axial partner. For a bounded spectrum
\footnote{See footnote 1 above.}, this expression reduces to
\be
\omega = \frac{\pi \varrho(i m)}{m}\  - 
\frac{\partial}{\partial m} \pi \varrho(i m)\ 
\ee
Below the critical temperature and as $m$ goes to zero,
the first term diverges (pion pole), while the second 
term approaches zero (plateau), provided that 
$\varrho (im)$ does not develop a low fractional dependence
on the sea quark mass (set equal to the valence one).
Above the critical temperature, the first term vanishes
for any value of the mass $m$,  which is followed by an infinite
jump in $\omega$ at  the critical temperature. For finite values of the mass 
$m$ this jump is smeared out. 

\begin{figure}[tbp]
\centerline{\epsfxsize=7.5cm \epsfbox{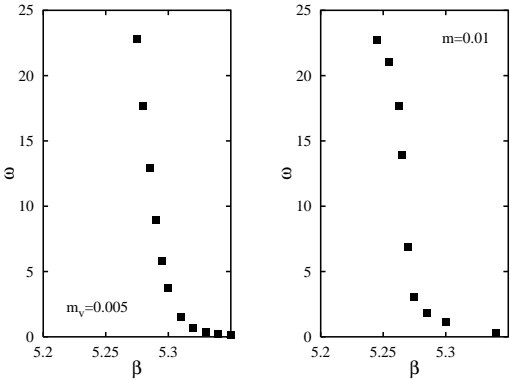}} 
\caption{(a) Results from the random matrix model for $m=0.005$ in units of 
$\Sigma=1$. (b) results for staggered fermions on a $16^3\times 4$ lattice 
with $ma= 0.01$ \protect\cite{CHRIST}.}
\label{fig3}
\end{figure}

Figure~\ref{fig3}a  shows the behavior for the Cardano solution with $m=0.005$ 
(0.5 MeV). Figure~\ref{fig3}b shows the same curve for $ma=0.01$ (6 MeV)
as obtained by Chandrasekharan and Christ on a $16^3\times 4$ lattice with 
staggered fermions \cite{CHRIST}. The similarity is striking, although the 
mass scales are different. The fact that in the random matrix model, $\varrho 
(im)$, does not develop a low fractional dependence on $m$ is traced back to
the structure of the thermal masses (T-regulated). We recall that the random 
matrix model considered here has no bearing on the axial-singlet anomaly. Its 
discussion in the context of random matrix models goes outside the scope of 
this work.

\vskip .5cm
{\bf 3.\,\,\,}
Wilson fermion formulations are more natural to use for three-flavor QCD. 
Recently Kalkreuter has performed detailed lattice simulations with Wilson 
fermions for two-color and two-flavor QCD \cite{KALKREUTER}. His results
are in qualitative agreement with random matrix theory \cite{JUREK,US}
for the bulk spectrum. The issue with Wilson fermions, however, 
is that the Wilson terms needed to eliminate the lattice doublers 
break explicitly chiral symmetry. It is however expected, that in the continuum
limit the breaking is mild enough (irrelevant operators) to allow for a 
full restoration of chiral symmetry, modulo mass terms. 

In terms of chiral random matrix models, the pertinent ensemble for 
investigating Wilson fermions for QCD is the gaussian unitary ensemble
\footnote{For QCD with two colors the relevant ensemble is the gaussian 
orthogonal ensemble \cite{JAC}.}, provided that the quarks are in the 
fundamental representation. In the presence of Wilson r-terms, the Dirac 
operator is only 
hermitean. It is not block-off-diagonal (chiral). Randomness, however, is 
sufficient to cause the spectral distribution
 to develop a non-vanishing value 
at zero virtuality and in the chiral limit. So, although the spectrum is not
paired, the distribution is even. This is easily understood if the diagonal 
disorder (r-terms) is not large enough \cite{US}. In fact, this can be easily 
investigated by considering a linear combination of two chiral random matrix 
models. One with off-diagonal randomness (chiral) plus one with diagonal
randomness (r-terms) \cite{US1}.

For three flavors, we define the hermitean Dirac operator
${\bf Q} = {\displaystyle \otimes_{f}} \gamma_5 (D\!\!\!\!/ +  m_f )$, 
with $f=u,d,s$. Here ${\bf Q}$ is the Dirac $Hamiltonian$ in five-dimensions. 
The Dirac matrix $\gamma_5$ plays the role of $\beta$-Dirac matrix
in (3+1) dimensions. 
On a cylinder, the pertinent random matrix model for
spatially constant Wilson modes is 
\be
{\bf Q}_W = \bigotimes_{f=u,d,s} \left(
          \left( \begin{array}{cc} m_f & \partial_{\tau}\\
                                   -\partial_{\tau} & - 
           m_f \end{array}\right) + {\bf R} \right)
\label{222}
\ee
where $\partial_{\tau}$ is the $\tau$-derivative on a cylinder of length 
$\beta= 1/T$, and ${\bf R}$ a $\tau$-independent random matrix
with a block structure specified by eq.(\ref{2a}) below. 
Equation~(\ref{222}) is defined
with anti-periodic boundary conditions. It is diagonal in frequency space
using 
\be
\psi (\tau ) = \sum_n e^{i\omega_n \tau} \psi_n 
\label{x1}
\ee
where $\omega_n = (2n+1) \pi T$ are the usual Matsubara frequencies.
Here~$\psi (\tau )$ is an $N$-dimensional vector for each of the three flavors:
up, down and strange. For each Matsubara frequency, and for each flavor
(\ref{222}) describes a set of $N$ fermions, with a hermitean Hamiltonian
\be
{\bf Q}_n^f = \left( \begin{array}{cc} m_f & i\omega_n \\
                                   -i\omega_n  & - m_f \end{array}\right)  +
                   \bigg(\ \mbox{\Large $\displaystyle R^f_n$}\ \,\bigg)
\label{2a}
\ee
This is a the sum of a deterministic and a random piece. Each entry in the 
first matrix is $N\times N$ valued. The random matrix is $2N\times 2N$ valued.
In terms of (\ref{2a}),
eq.~(\ref{222}) is block-diagonal in frequency and in flavor space,
\be
{\bf Q}_W = \bigotimes_{f=u,d,s} \bigotimes_{n=-\infty}^{+\infty} {\bf Q}_n^f 
\label{x2}
\ee

The random distributions for each Matsubara mode are decoupled,
each with a weight
\be 
P( { R_n })=\frac{1}{Z} \exp [-2N {\rm Tr} V({ R_n})]
\label{4}
\ee
This is a direct consequence of our assumption that  ${\bf R}$ in 
(\ref{222}) is $\tau$-independent. If we were to relax this condition, then the
random matrix may cause the Matsubara modes to couple \cite{US1}.
In this paper, $V$ will be chosen gaussian, although
 polynomial weights are also possible \cite{BREZIN}. The resolvent of 
(\ref{x2}) is
\be 
G(z)= \frac{1}{2NN_fN_*}\left< {\rm Tr} \frac{1}{z-{\bf Q}} \right>
\label{3}
\ee
The averaging is carried with the distribution~(\ref{4}).
In terms of (\ref{3}) the spectral distribution associated to (\ref{222}) 
\be
\nu (\lambda ) = \frac 1{2N_f NN_*} \sum_{n=-\infty}^{+\infty}
\sum_{f=u,d,s}
{\rm Tr}_{2N} \, \delta ( \lambda - {\bf Q}_n^f )
\label{x3}
\ee
is related to the discontinuity of $G(z)$ through the real axis
\be 
\nu(\lambda )=-\frac{1}{\pi}
\,{\rm Im} G(z=\lambda +i 0)
\label{5}
\ee
Equation~(\ref{x3}) involves the sum over a large number of Matsubara modes,
with $N_*= \sum_n$. 

Using either the law of addition of random matrices \cite{US,ZEE} or
diagrammatic techniques \cite{BREZINZEE}, it follows that in large $N$
the resolvent $G(z)$
is 
\be
G(z)=\frac{1}{N_f N_*}\sum_{n,f} G_n^f(z)
\label{eachres}
\ee
where each of the $G_n^f(z)$ satisfying Pastur's equation \cite{PASTUR} 
\be
G_n^f(z)= \frac 1{2}
\left(
\frac{1}{z-G_n^f(z)- {\bf M}_n^f}+\frac{1}{z-G_n^f(z)+ {\bf M}_n^f}\right)
\label{6}
\ee
is a 2$\times$2 diagonal matrix, with $M_n^f (T) =\sqrt{ m_f^2 + \omega_n^2}$.

At high temperature, only few modes contribute to (\ref{eachres}), $i.e.$ 
$\omega_0 = \omega_{-1} = \pi T$.
Taking
$m_u\sim m_d$, the problem reduces to
\be
G(z) &=& \frac{\alpha}{2} G^u(z) + \frac{(1-\alpha)}{2} G^s(z)
\label{greentwo}
\ee
with $\alpha = 1-1/N_f$, $N_f=3$ and $G^f(z)$ defined by~(\ref{6}).
(\ref{greentwo}) is a combination of three third order algebraic equation
(Cardano class) \cite{US}, for the resolvent $G(z)$.

The spectral density follows from the discontinuity 
of $G(z)$ along the imaginary axis (\ref{5}). In the chiral limit, the 
discontinuity is related to the chiral condensate through
the Banks-Casher argument,
\be 
<\bar{q} q> = {\rm Im} \,\,G(z = i 0 )\,\,\,.
\label{10}
\ee
Although (\ref{greentwo}) was derived in the high 
temperature limit, we expect it to be rather accurate near the critical points
because of universality. Indeed, at the critical point and in large $N$ only
symmetry is relevant \cite{MACIEKII}. Below, we will provide an analysis of
the spectral distribution associated to (\ref{greentwo}).

\vskip .5cm
{\bf 4.\,\,\,}
The explicit solution of the third order equation is available in 
algebraic form, and was discussed in \cite{US}. The combination of several
third order equations allows for a rich phase structure. In the case of two 
light and one heavy flavors, the spectrum should exhibit at most four phases 
$P_1, P_2, P_3$ and $P_4$. The support of the spectral function may consists
of one, two, three or four disconnected arcs, depending on the choice of the 
external parameters $m_u$, $m_s$, and $T$.

\begin{figure}[tbp]
\centerline{\epsfxsize=7.5cm \epsfbox{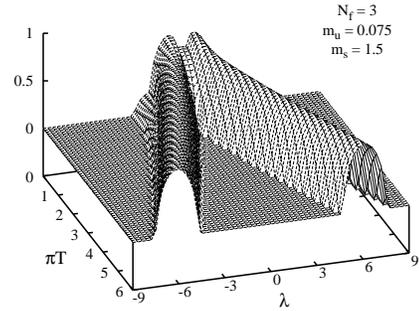}}
\caption{Spectral function for Wilson fermions with $m_u=m_d=0.075$ and 
$m_s=1.5$.}
\label{figd1}
\end{figure}

Figure~\ref{figd1}
shows the behavior of the spectral function for the physical choice of 
$m_u=m_d=0.075, m_s=1.5$. All dimensionfull quantities are measured in units 
of $\Sigma =1$. In physical units, $m_u=m_d=7.5$MeV and $m_s=150$MeV.
At  zero temperature, the system is in the $P_1$ phase, and the value at 
zero virtuality ($\lambda=0$) corresponds to a non-zero condensate. 
The mass of the strange quark is 20 times higher, therefore the spectrum
feels it by developing two symmetric shoulders. The mass is not high enough, 
however, to provide the decoupling of the strange quark ($P_3$ phase),
with symmetric and disconnected humps. The effect of the temperature 
is to quench the maximum at the origin. At the critical temperature $T_*$,
the quarks localize, resulting into a transition from $P_1\rightarrow P_2$. 
For increasing $T$, the thermal quark modes contribute democratically to
the spectral distribution, $M_u=\sqrt{m_u^2 +\pi^2 T^2}\rightarrow \pi T$, 
$M_s=\sqrt{m_s^2 +\pi^2 T^2}\rightarrow \pi T$.

\begin{figure}[tbp]
\centerline{\epsfxsize=7.5cm \epsfbox{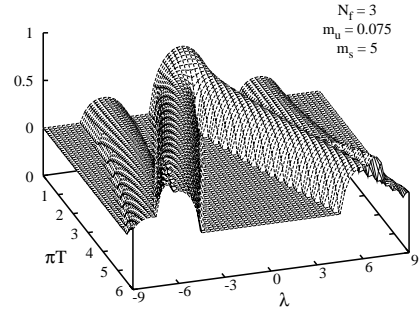}}
\caption{Spectral function for Wilson fermions with $m_u=m_d=0.075$ and 
$m_s=5$.}
\label{figd2}
\end{figure}

Figure~\ref{figd2} shows the same spectral distribution as Fig.~\ref{figd1},
 but this time for 
a large strange quark mass $m_s=5$, or $m_s=500$ MeV in physical units. 
The two additional humps at $T=0$ are just the delocalized strange quarks.
An analysis of Cardano's solutions show that the critical value for which this 
happens is $2m_s^2 = 11+ \sqrt{125}$ for $m_u=0$. Here, 
the initial spectrum is 
in the $P_3$ phase. The effects of the temperature is to cause localization of 
the light quarks, first transition, from $P_3$ to $P_4$, resulting in the 
vanishing of the density of states at zero virtuality.  This is followed by a 
second transition at higher temperature $T_{**}$
from $P_4$ to $P_2$ in which all the three flavors become thermally heavy.

\begin{figure}[tbp]
\centerline{\epsfxsize=7.5cm \epsfbox{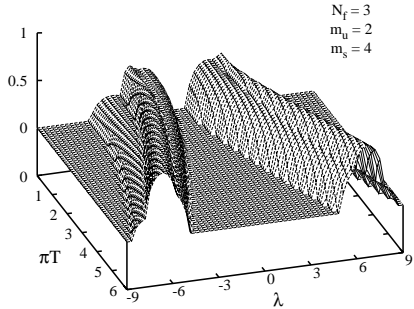}}
\caption{Spectral function for Wilson fermions with $m_u=m_d=2$ and $m_s=4$.}
\label{figd3}
\end{figure}

Figure~\ref{figd3}
 shows the same spectral distribution for three large but comparable 
masses, $m_u=m_d=2$ and $m_s=4$. There is no phase transition, the system is 
always in the heavy (localized) phase $P_2$. There is no accumulation of quark 
states at zero virtuality. The thermal effects cause the localization to shift 
from the Compton wavelength to the thermal wavelength.

\begin{figure}[tbp]
\centerline{\epsfxsize=7.5cm \epsfbox{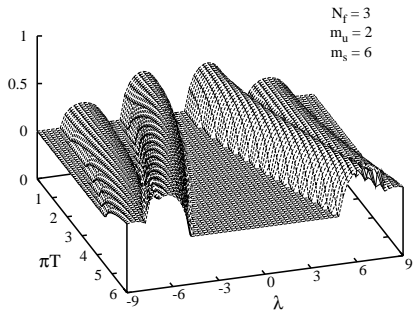}}
\caption{Spectral function for Wilson fermions with $m_u=m_d=2$ and $m_s=6$.}
\label{figd4}
\end{figure}

Figure~\ref{figd4} shows yet another 
choice of heavy quark masses $m_u=m_d= 2$ and 
$m_s=6$, for which the spectral distribution undergoes a phase transition from 
$P_4$ to $P_2$. At low temperature the 
quarks are massive but distinct, at high 
temperature the distinction is washed out by temperature.

The spectra discussed above for the random matrix model,
allow for an assessment of the critical lines (phase diagram)
in terms of the end-points of the spectral distribution. The equations
for the end-points are given by a polynomial equation that is always one degree
lower than the one following from Pastur's equation. This is best described 
using Blue's functions \cite{ZEE}, i.e. functional inverses of Green's  
functions. Generically, the Blue's function associated to the Green's function
$G(z)$ is given by
\be 
 B(G(z)) =z
\label{blue}
\ee
The substitution $z \rightarrow B(z)$ and the use of  (\ref{blue}) 
in Pastur's equation (\ref{greentwo}) yield the pertinent equation for
the Blue's functions. Since the Green's functions behave as $\sqrt{A-z}$ 
when $z$ approaches  the endpoint $A$, then 
\be
\frac{dG}{dz}|_{end-point} = \infty \,\,\,\leftrightarrow \,\,\,\,
\frac{dB}{dz}|_{end-point} = 0
\label{boundary}
\ee
For our case, this yields a quadratic equation \cite{US}. The end points are 
explicitly given by
\be
A_f^{\pm} = \frac{1}{\sqrt{2}} \sqrt{1 + 2 M_f^2 \pm%
  \sqrt{1 + 8 M_f^2}} 
  \frac{3 \pm\sqrt{1 + 8 M_f^2}}{1 \pm\sqrt{1 + 8 M_f^2}}
\label{4endpoints}
\ee
for all masses and temperatures.
In the case of Figure~\ref{figd2} and \ref{figd4}, the transition from $P_4$
to $P_2$ is defined as $A_u^+ = A_s^-$. In the high mass (temperature)
limit this condition is equivalent to $M_s(T_{**}) - M_u(T_{**}) = 2\sqrt{2}$.
At zero virtuality the spectral distribution vanishes for 
$\pi T_* = \sqrt{1 - m_u^2}$.

{\bf 5.\,\,\,\,\,\,}
We have shown that the chiral gaussian unitary ensemble for two massive flavors
and finite temperature, allows for a quantitative description of the recent simulations
by Chandrasekharan and Christ using staggered fermions. The agreement with the 
random matrix model at $\beta =5.34$ and higher suggests that the fit to the 
lattice data with $\delta =5/3$ may be underestimating the critical exponent
at $\beta= 5.275$ in comparison with the mean-field result by about a factor of 
two. We have also provided a qualitative understanding of the lattice 
anomalous symmetry breaking effect in a chiral random matrix model 
without the axial-singlet anomaly, albeit with a small sea quark mass.

We have suggested that in the chiral limit the gaussian unitary ensemble is 
the pertinent random ensemble for investigating Wilson fermions in QCD.
We have explicitly shown that the problem of three flavor QCD
in the Wilson formulation is amenable to a random plus deterministic problem 
for few Matsubara modes. We have used arguments at high temperature to unravel 
a number of properties of the spectral distribution for the hermitean 
Dirac operator (Hamiltonian in five dimensions). QCD with two light and one 
heavy flavor  exhibits a spontaneously broken phase at low temperature, and 
a localized phase at high temperature. The localization is determined by the 
thermal wavelength. For QCD with massive flavors, the localization is a 
competitive effect between the thermal wavelength and the Compton wavelength.

We have explicitly used the end-points of the Dirac spectrum to construct 
the critical lines for massive QCD at finite temperature. The critical lines 
(phase diagrams) were shown to follow from a quadratic equation following from 
the Blue's function. Given the qualitative agreement noted between our results, 
and the staggered fermion calculations, we look forward to a detailed 
comparison with future simulations using Wilson fermions for full QCD. 
We note, however, that  our analysis does not account for the 
axial-singlet anomaly. Its role in the context of random matrix models
will be discussed elsewhere. 

The phase transitions discussed here  are all mean-field in character. 
This is unavoidable in the large $N$ analysis performed 
here. The quantitative similarities between the Columbia results at finite 
temperature and the random matrix results, suggest that the phase transition 
in quarkish QCD may be characterized by a narrow Ginzburg window 
(perhaps of order $1/N$), thereby asserting the lore of mean-field exponents.
Within the Ginzburg window  scaling arguments and universality should hold
\cite{PISARSKI}.

The spectral distributions we have discussed here are of course amenable
to finite $N$ corrections. This 
could be discussed either in the context of the 
corrections to Pastur's equations or using diagrammatic techniques. Also, 
specific level correlations can be discussed either in the staggered 
formulation, or the Wilson one, both for massive QCD and finite temperature. 
These and other issues will be discussed elsewhere.

\vglue 0.6cm
{\bf \noindent  Acknowledgments \hfil}
\vglue 0.4cm
This work was supported in part  by the US DOE grant DE-FG-88ER40388,
by the Polish Government Project (KBN) grant 2P03B19609 and by 
the Hungarian Research Foundation OTKA.

\vskip 1cm
\setlength{\baselineskip}{15pt}


\begin{thebibliography}{50}

\bibitem{KANAYA}
K. Kanaya, "Finite Temperature QCD on the Lattice", Nucl. Phys. {\bf B} (1996).

\bibitem{IWASAKI}
Y. Iwasaki, Nucl. Phys. (Proc. Suppl.) {\bf B42} (1995) 96.

\bibitem{DETAR}
C. DeTar, Nucl. Phys. (Proc. Suppl.) {\bf B42} (1995) 73.


\bibitem{BOOK}
M.A. Nowak, M. Rho and I. Zahed, {\it Chiral Nuclear Dynamics},
 World-Scientific (1996), in print


\bibitem{PISARSKI}
R.D. Pisarski and F. Wilczek, Phys. Rev. {\bf D29} (1984) 338.


\bibitem{RANDOMOTHERS}
See e.g., C.E. Porter, {\it Statistical Theories of Spectra: Fluctuations},
Academic Press, New York, 1965;
M.L. Mehta, {\it Random Matrices}, Academic Press, New York, 1991. 


\bibitem{RANDOMQCD}
E.V. Shuryak and J.J.M. Verbaarschot, Nucl. Phys. {\bf A560} (1993) 306.
J.J.M. Verbaarschot and I. Zahed, Phys. Rev. Lett. {\bf 70}
(1993) 3852.


\bibitem{BEFORE}
A.D. Jackson and J.J.M. Verbaarschot, `` A random matrix model for chiral 
symmetry breaking'', SUNY-NTG-95/26, eprint hep-ph/9509324.



\bibitem{STEPHANOV}
M.A. Stephanov, ``Chiral symmetry at finite T, the phase of Polyakov loop
and the spectrum of the Dirac operator'', eprint hep-lat/9601001.


\bibitem{US}
M.A. Nowak, G. Papp and I. Zahed, ``QCD-inspired spectra from Blue's 
functions", e-print hep-ph/9603348.


\bibitem{CHRIST}
S. Chandrasekharan and N. Christ, ``Dirac Spectrum, Axial Anomaly and the QCD
Chiral Phase Transition'', eprint hep-lat/9509095.



\bibitem{REDUCTION}
T.H. Hansson and I. Zahed, Nucl. Phys. {\bf B374} (1992) 117;
G.E. Brown et al., Phys. Rev. {\bf D45} (1992) 3169;
T.H. Hansson, M. Sporre and I. Zahed, Nucl. Phys. {\bf B427} (1994) 545. 


\bibitem{MICRO}
J.J.M. Verbaarschot, Phys. Lett. {\bf B368} (1996) 137.


\bibitem{KARSCH}
F. Karsch and E. Laermann, ``Susceptibilities, the specific heat and a cumulant 
in two-flavor QCD", BI-TP 94-29.

\bibitem{KALKREUTER}
T. Kalkreuter, ``Numerical analysis of the spectrum of the Dirac 
operator in four-dimensional $SU(2)$ gauge fields'', eprint hep-lat/9511009.

\bibitem{JUREK}
J. Jurkiewicz, M.A. Nowak and I. Zahed, ``Dirac spectrum in QCD and quark 
masses", e-print hep-ph/9603308.


\bibitem{JAC}
J.J.M. Verbaarschot, in Continuous Advances in QCD, Ed. A.V. Smilga, World 
Scientific (1994).


\bibitem{US1}
M.A. Nowak, G. Papp and I. Zahed, in preparation.


\bibitem{BREZIN}
E. Brezin, S. Hikami and A. Zee, "Oscillating density of states near zero 
energy 
for matrices made of blocks with possible application to the random flux 
problem", eprint cond-mat/9511104.


\bibitem{ZEE}
A. Zee, ``Law of addition in random matrix theory'', eprint cond-mat/9602146.


\bibitem{BREZINZEE}
E. Brezin and A. Zee, Phys. Rev. {\bf E49} (1994) 2588.


\bibitem{PASTUR}
L.A. Pastur, Theor. Mat. Phys. (USSR) {\bf 10} (1972) 67.

\bibitem{MACIEKII}
M.A. Nowak and I. Zahed, in preparation.



\end{thebibliography}
\end{document}